\documentclass[10pt, letterpaper]{IEEEtran}

\usepackage{ifpdf}
\usepackage{amsmath,amssymb,amsfonts,bm}
\usepackage{graphicx}
\usepackage{tabularx}
\usepackage{enumitem}
\usepackage{mathtools}
\usepackage{float}
\usepackage{textcomp}
\usepackage{moresize}
\usepackage{multirow}
\usepackage[ruled,linesnumbered,noend]{algorithm2e}
\usepackage{algpseudocode}
\usepackage{breqn}
\usepackage{subcaption}
\usepackage{hyperref}
\usepackage{balance}
\usepackage{ tipa }
\usepackage{xcolor}
\usepackage{xfrac}
\usepackage{footnote}
\makesavenoteenv{algorithm}
\DeclareMathOperator*{\argmax}{argmax}

\usepackage{fancyhdr}
\usepackage{soul}

\newcommand{\mco}{\mathcal{O}}

\allowdisplaybreaks

\usepackage{amsthm}

\makeatletter
\newcommand{\vast}{\bBigg@{4}}

\newcommand{\Vast}{\bBigg@{5}}

\begin{document}

\title{Learning-Based Two-Way Communications: Algorithmic Framework and Comparative Analysis}


\author{\IEEEauthorblockN{David R. Nickel,~\IEEEmembership{Graduate Student Member,~IEEE},
Anindya Bijoy Das,~\IEEEmembership{Member,~IEEE}, David J. Love,~\IEEEmembership{Fellow,~IEEE} and
Christopher G. Brinton,~\IEEEmembership{Senior Member,~IEEE}}

\thanks{\IEEEauthorblockA{\noindent David R. Nickel, David J. Love, and Christopher G. Brinton are with the Elmore Family School of ECE, Purdue University, West Lafayette, IN, USA (\{dnickel, djlove, cgb\}@purdue.edu). Corresponding author: David R. Nickel.\\\indent Anindya Bijoy Das is with Deptartment of ECE, The University of Akron, Akron, OH, USA (adas@uakron.edu).}\\\indent This work was supported in part by the Office of Naval Research (grant N00014-21-1-2472), RTX Corporation, and the NDSEG Fellowship Program.}}

\maketitle
\begin{abstract}
Machine learning (ML)-based feedback channel coding has garnered significant research interest in the past few years. However, there has been limited research exploring ML approaches in the so-called ``two-way" setting where two users jointly encode messages and feedback over a shared channel. In this work, we present a general architecture for ML-based two-way feedback coding, and show how several popular one-way schemes can be converted to the two-way setting through our algorithmic framework. We compare such schemes against one-way counterparts, revealing error-rate benefits of ML-based two-way coding in certain signal-to-noise ratio (SNR) regimes. We then analyze the tradeoffs between error performance and computational overhead for three state-of-the-art neural network coding models instantiated in the two-way paradigm.
\end{abstract}




\section{Introduction}\label{sec:intro}
In modern communication systems, bidirectional exchange of information can be critical, as in, for example, communications between a base station and a satellite. Such a system is referred to as a ``two-way channel" \cite{shannon_twc_1961}. Focus has been dedicated to settings in which additive white Gaussian noise (AWGN) is added independently in both directions, called Gaussian two-way channels (GTWCs). A key early result on two-way coding demonstrated that when noise is independently introduced in both directions, the optimal channel capacity is attained by treating each direction as a separate one-way channel \cite{han_independentgtwc_1984}. More recently, \cite{palbaus_errorexponent_2021} established potential benefit for error exponents in cooperative GTWCs over non-cooperative cases.

Two-way (TW) feedback coding is itself an extension of the Gaussian one-way (OW) feedback channel model (GOWC), pioneered in \cite{sk_gowc_1966}. While this initial OW scheme was effective in cases of no feedback noise, it would be several decades before the development of GOWC schemes which were effective with noisy feedback \cite{chance_linearcoding_2011,yishai_fbkcode_2017}. More recently, works on GOWCs began employing machine learning (ML) to perform the encoding and decoding tasks. As discussed below, these ML techniques have led to significant improvement in performance metrics in certain signal-to-noise ratio regimes. 

\subsection{Non-Linear One-way Coding}
Due to the time-series-like nature of the feedback information, early ML-based schemes such as DeepCode \cite{hkim_deepcode_2020} utilized recurrent neural networks (RNNs) for encoding and decoding messages, with focus later shifting to transformer \cite{vaswani_xformer_2017} models. 
Contemporary advances in one-way coding have produced results surpassing those of earlier ML schemes while also being able to operate at any coding rate. The first, Generalized Block Attention Feedback Coding (GBAF) \cite{ozfatura_gbaf_2022} saw orders of magnitude of improvement on block error-rate (BLER) over prior RNN-based schemes and its predecessor, AttentionCode \cite{shao_attentioncode_2023}. Further progress in transformer-based architectures came via Block Attention Active Feedback Codes \cite{ozfatura_baaf_2023}, which took the original structure of GBAF and used another encoder at the receiver to transmit feedback symbols, rather than just feeding back the received symbols themselves.

Another major advance came from LightCode (LC) \cite{ankireddy_lightcode_2024}, which found that in certain use-cases, the attention mechanisms in GBAF provide marginal benefit in system performance. Building upon GBAF's notion of a feature extractor, \cite{ankireddy_lightcode_2024} formulated a system with only a small feed-forward network with a skip connection. LightCode's encoder/decoder architecture achieved comparable performance to GBAF with only 10\% of the trainable weights. In reducing the number of weights, LC is more time- and space-efficient than its transformer- and RNN-based counterparts. Lastly, the authors of \cite{jkim_robustcode_2023} developed an RNN-based model, RobustCode, featuring an attention module at its decoder's output. RobustCode is the basis for the two-way RNN-based scheme discussed below.

\begin{figure*}[ht]
    \centering
    \includegraphics[height=4.3cm,width=.8\textwidth]{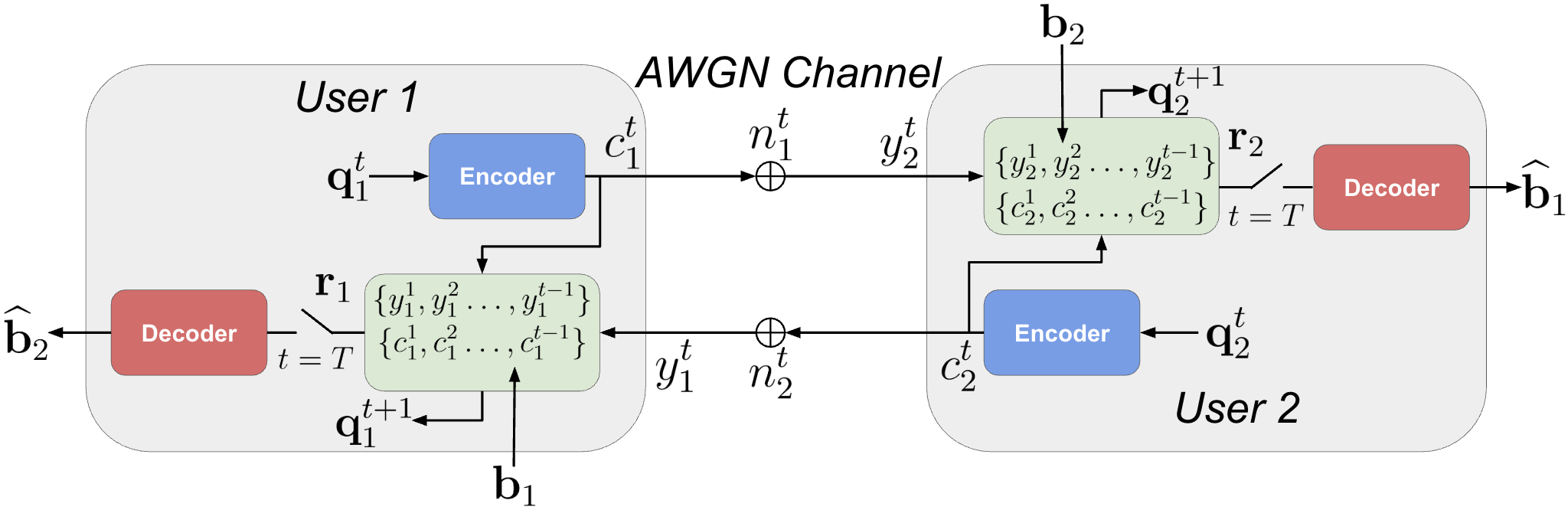}
    \caption{\small System model for two-way feedback coding. At time $t$, user $i\in\{1,2\}$ encodes transmit-side knowledge vector $\bm{q}_{i}^{t}$ into transmit symbol $c_i^t$ and receives $y_i^t$ over an AWGN channel. $c_i^t$ and $y_i^t$ are stored to update the knowledge vector into $\bm{q}_i^{t+1}$. After $T$ transmissions, receive-side knowledge vector $\bm{r}_i=[\bm{b}_i,\bm{y}_i]^{\intercal}$ is passed through the decoder to generate decoded message $\widehat{\bm{b}}_j$, $j\neq i$.}\label{fig:sys_diag_generic}
\end{figure*}

\subsection{Two-way Coding Schemes}
Despite considerable progress in the development of one-way codes, as well as some recent explorations of broadcast channels \cite{malayter2024deeplearningaidedbroadcast,zhou2025learnedcodesbroadcastchannels}, there exists relatively little work examining feedback coding in two-way systems. Recently, \cite{jkim_rnntwc_2023} was the first to explore finite block-length two-way coding, presenting both a linear system developed akin to \cite{chance_linearcoding_2011} and an extension of RobustCode to the two-way case (TWRNN) which is trained to minimize the sum-BLER of the users. The work tackles three difficulties of GTWCs: (1) the coupling of the two users' encoding processes, (2) the need for decoders capable of parsing through this coupling, and (3) the need for effective power allocation within a given power constraint. TWRNN displays robustness by performing well in low SNR regimes, while the linear scheme excels at high SNR. 


Despite the considerable work put forward in \cite{jkim_rnntwc_2023}, many facets of two-way coding are yet to be explored. In seeing the success of both GBAF and LC architectures, we find a natural route of inquiry in assessing their viability in the two-way paradigm as alternatives to RNN-based schemes.


\subsection{Summary of Contributions}\label{ssec:contributions}
\begin{itemize}
    \item We present an algorithmic framework for optimizing sum block error rate in non-linear encoder/decoder systems \textcolor{black}{with a generalized input space which allows the use of more of the available past information}. With this architecture, we develop new ML-based two-way coding schemes, Two-way Block Attention Feedback (TWBAF) and Two-way LightCode (TWLC), based on the one-way models found in \cite{ozfatura_gbaf_2022,ankireddy_lightcode_2024}.
    \item \textcolor{black}{We assess the tradeoffs between OW and TW coding schemes by comparing (i) TWLC against a novel active feedback-enabled OW variant which we develop, Active LightCode, and (ii) the linear two-way scheme of \cite{jkim_rnntwc_2023} against the inner code of \cite{chance_linearcoding_2011}. Results show distinct improvements can be wrought from the TW paradigm, though this is both SNR-regime- and model-dependent.}
    \item We investigate the performance-complexity tradeoffs between different ML-based two-way codes through numerical experiments comparing BLER and analysis of floating point operations (FLOPS) executed by each. \textcolor{black}{In doing so, we provide insights into the efficacy of different model types for future works in two-way system design.}
\end{itemize}

\section{\textcolor{black}{System Model and Optimization}}\label{sec:system_model}
We consider two users, $i=1,2$, each aiming to transmit bitstreams $\bm{b}_{i}\in\{0,1\}^{K_i}$ at rate $R_i=K_i/T$. Let $t=1,\dots,T$ be the channel use indices, and let $c_{i}^{t}\in\mathbb{R}$ be the symbol sent by user $i$ at time $t$, as generated by a transmit-side neural network-based encoder; each user also possesses a receive-side ML-enabled decoder. The received symbol at user $i$ is then
\begin{equation}
    y_{i}^{t}=c_{j}^{t}+n_{j}^{t},~j\neq i,~i,j\in\{1,2\},
\end{equation}
\noindent where $n_{j}^{t}\sim\mathcal{N}(0,\sigma_{j}^{2})$ is i.i.d. $\forall t$. We collect these as $\bm{n}_{i}=[n_{i}^{1},\dots,n_{i}^{T}]$, $\bm{c}_{i}=[c_{i}^{1},\dots,c_{i}^{T}]$, and $\bm{y}_{i}=[y_{i}^{1},\dots,y_{i}^{T}]$ and subject each user to a power constraint given by 

\begin{equation}\label{eq:pwr_constraint}
    \mathbb{E}_{\bm{b}_{i}, \bm{n}_{1},\bm{n}_{2}}\left[\sum\nolimits_{t=1}^{T}c_{i}^{t}\right]\leq P_iT,
\end{equation}

\noindent where $P_i$ is a constant for power. Lastly, we denote the signal-to-noise ratio (SNR) from the transmitter of user $i$ to the receiver of user $j$ as $SNR_i$.

\subsection{ML Model Inputs and Outputs}
Similar to \cite{ozfatura_gbaf_2022}, we define transmit-side knowledge vectors $\bm{q}_{i}^{t}\in\mathbb{R}^{K_i+2(T-1)}$ as the encoders' inputs. Define the overall functional representation of generating transmit symbols at the encoder at time $t$ as $c_{i}^{t}=E_i(\bm{q}_{i}^{t})$,\footnote{In \cite{shannon_twc_1961}, Shannon considers a sequence of time-indexed encoding functions. It is more convenient for us to consider a single encoding function which is implicitly time-indexed by the knowledge vectors to simplify the presentation of the ML-based optimization scheme.} which is power-constrained under the trainable power reallocation scheme from \cite{jkim_rnntwc_2023}. The knowledge vectors are instantiated at $t=1$ as:
\begin{equation}\label{eq:know_vec_inst}
    \bm{q}_{i}^{1}=[\bm{b}_{i},\bm{0}_{2(T-1)}]^{\intercal}.
\end{equation}
$\bm{0}_{2(T-1)}$ denotes a length-${2(T-1)}$ padding vector that will be replaced by previously sent symbols, $c_{i}^{t'<t}$, and by previously received information $\tilde{y}_{i}^{t'<t}$, which equals either $y_{i}^{t'<t}$ or $y_{i}^{t'<t}-c_{i}^{t'<t}$, over subsequent channel uses.\footnote{For simplicity of notation, we henceforth suppress the tilde over $y$.} At $t=2$, the knowledge vectors are updated as
\begin{equation}
    \bm{q}_{i}^{2}=[\bm{b}_{i},c_{i}^{1}, \bm{0}_{T-2},{y}_{i}^{1},\bm{0}_{T-2}]^{\intercal},
\end{equation}
continuing iteratively until the final channel use at time $T$:
\begin{equation}\label{eq:know_vec_final}
    \bm{q}_{i}^{T}=[\bm{b}_{i}, c_{i}^{1}, c_{i}^{2},\dots,c_{i}^{T-1}, y_{i}^{1},y_{i}^{2},\dots,y_{i}^{T-1}]^{\intercal}.
\end{equation}

\textcolor{black}{After $T$ channel uses, we form receive-side knowledge vectors as $\bm{r}_{i}=[\bm{b}_{i}, \bm{y}_{i}, \textcolor{black}{\bm{c}_i}]^{\intercal}$. Let $D_{i}=\mathbf{O}_i\circ \argmax(\widehat{D}_i)$} denote the functional representation of user $i$'s decoder.\footnote{Decoder $D_i$ is housed at user $j\neq i$.} $\widehat{D}_i$ is the output of the receive-side neural network, and $\mathbf{O}_i$ is given shortly. Let $\widehat{\bm{d}}_{i}=\widehat{D}_{i}(\bm{r}_{j})\in\Delta^{2^{K_i}}, i\neq j$, where $\Delta$ is the unit-simplex. Let $\mathbf{O}_i:\bm{k}\rightarrow\{0,1\}^{K_i} $ be a bijection mapping $\bm{k}$, a one-hot vector with non-zero entry at index $k$, to a length-$K_i$ binary bitstream. The estimated bitstream for user $i$ is then 
\begin{equation}\label{eq:est_bitstream}
\widehat{\bm{b}}_{i}=\mathbf{O}_i(\argmax{\widehat{\bm{d}}_{i}}).
\end{equation}
A diagram of the overall transmission and decoding scheme is presented in Fig. \ref{fig:sys_diag_generic}.

\subsection{Optimization of Encoders and Decoders}
We consider the metric of minimizing block error rate (BLER), which is the ratio of the number of incorrect estimated bitstreams to the total number of bitstreams sent:

\begin{equation}
    \mathcal{E}_{i}(E_i,D_i)=\mathsf{Pr}\left[\{\bm{b}_{i}\neq\widehat{\bm{b}}_{i}\}\right],~i\in\{1,2\}.
\end{equation}
The goal of the system is therefore to minimize the sum BLER of the users by jointly optimizing over their encoders and decoders, subject to \eqref{eq:pwr_constraint}, i.e.,
\begin{align}\label{eq:opt_prob}
    \min_{E_{1},E_{2},D_{1},D_{2}}~~~&\mathcal{E}_{1}(E_{1},D_{1})+\mathcal{E}_{2}(E_{2},D_{2})~~~\text{s.t.}~\eqref{eq:pwr_constraint}.
\end{align}

Since $\bm{d}_{i}$ is mapped to a unique sequence of ones and zeros via $\mathbf{O}_i$, we have that 
\begin{equation}
    \mathsf{Pr}\left[\{\bm{b}_{i}\neq\widehat{\bm{b}}_{i}\}\right]=\mathsf{Pr}\left[\{\mathbf{O}_i^{-1}\left(\bm{b}_{i}\right)\neq \widehat{\bm{d}}_{i}\}\right],~i=1,2.
\end{equation}

The optimization problem in \eqref{eq:opt_prob} can therefore be considered a classification problem over $2^{K_i}$ classes, for which we adopt cross-entropy (CE) loss. Let $\bm{d}_{i}=\mathbf{O}_i^{-1}\left(\bm{b}_{i}\right)$, and parameterize user $i$'s models by $\theta_i$. The per-user and overall loss are then
\begin{gather}\label{eq:loss}
\mathcal{L}_{\theta_i}(\{\bm{d}_{i},\widehat{\bm{d}}_{i}\})=-\sum\nolimits_{\ell=1}^{2^{K_i}}d_{i}[\ell]\log\left(\widehat{d}_{i}[\ell]\right)\\
\text{and}~\mathcal{L} = \mathcal{L}_{\theta_1}+\mathcal{L}_{\theta_2}. 
\end{gather}
Lastly, let
\begin{equation}\label{eq:know_vec_update}
    \bm{q}_{i}^{t+1}=U(\bm{q}_{i}^{t}, \bm{y}_{i}^{t}, \bm{c}_{i}^{t})
\end{equation}
\noindent denote the function which updates the transmit-side knowledge vectors using \eqref{eq:know_vec_inst}-\eqref{eq:know_vec_final}. The overall GTWC optimization scheme is summarized for a single bitstream in Alg. \ref{alg:training}.

\section{\textcolor{black}{Channel Coding Architectures}}\label{sec:architectures}
\begin{figure}[t]
    \centering
    \includegraphics[width=\linewidth]{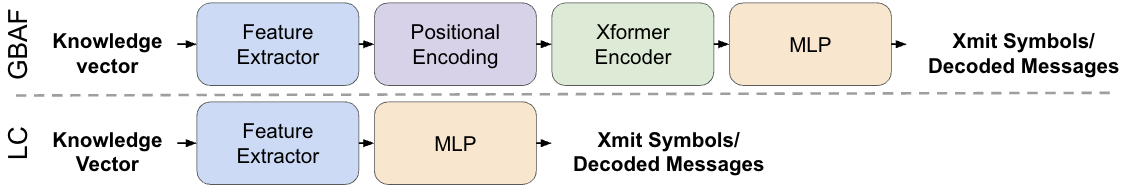}
    \caption{\small Block diagrams for GBAF and LC. The feature extractors (FE) and multi-layer perceptrons (MLPs) are fully-connected feed-forward networks; LC contains a skip-connection from its FE's first layer to its last.
    The positional encoding and transformer (Xformer) encoder block follow the standard implementations in \cite{vaswani_xformer_2017}. Model layer specifications for TWLC and TWBAF follow roughly those outlined in \cite{ankireddy_lightcode_2024, ozfatura_gbaf_2022}, with modifications as indicated in Sec. \ref{sec:architectures}.}\label{fig:lc_gbaf_blocks}
\end{figure}

Using the framework developed above, we present three new ML-based channel coding architectures: a OW variant of LC leveraging active feedback and two TW systems based on LC and GBAF. Block diagrams for LC and GBAF encoders/decoders can be found in Fig. \ref{fig:lc_gbaf_blocks}.

\noindent\textbf{TW LightCode-Based System (TWLC)}: We begin with the overall model shape described in \cite{ankireddy_lightcode_2024} (Figs. 12 \& 13 therein). In adapting it to a two-way system, we double the number of neurons per hidden layer with the goal of making the model more competitive in the BLER metric \textcolor{black}{while keeping the model relatively small as intended}. Additionally, layer normalization is added after the last layer of the feature extractor \textcolor{black}{for stability. While TWLC was bested in BLER by some other models (Fig. \ref{fig:results_all}), its training was by far the most robust, experiencing neither power-allocation issues nor training instability seen by TWRNN and TWBAF (below). Its smaller size also makes it more amenable to operation on smaller devices.} 

\begin{algorithm}[ht!]
\small
\caption{GTWC Training for a Single Bitstream}\label{alg:training}
\textbf{Input}: Bitstream length $K_i$, block-length $T$, noise power $\sigma_{i}^{2}$, power constraint $P_{i}$, model parameters $\theta_i$ (for $i=1,2$).\\
Generate bitstreams: $\bm{b}_{i}\in\{0,1\}^{K_i}$.\\
Make knowledge vectors: $\bm{q}_{i}^{1}\gets[\bm{b}_{i},\bm{0}^{2(T-1)}]$\\
\For{$t=1,...,T$}{
    Make transmit symbols, ${c}_{i}^{t}\gets E_{i}(\bm{q}_{i}^{t})$, then transmit.\\
    Generate noise ${n}_{i}^{t}$ $\&$ ${n}_{j}^{t}$. Store received symbols, ${y}_{j}^{t}={c}_{i}^{t}+{n}_{i}^{t}$.\\
    \If{$t<T$}{
        Update $\bm{q}_{i}^{t+1}$ using \eqref{eq:know_vec_update}.
    }
}
Make received knowledge vector: $\bm{r}_{i}\gets[\bm{b}_{i},\bm{y}_{i},\bm{c}_{i}]$\\
Decode symbols: $\widehat{\bm{d}}_{j}\gets \widehat{D}_{j}(\bm{r}_{i})$.\\
Calculate loss using \eqref{eq:loss}.\\
Back-propagate. Update gradient scheduler (if needed).

\textbf{Output}: Updated GTWC model parameters $\theta_i$.
\end{algorithm}

\noindent\textbf{OW Active Feedback LightCode System (ALC)}:
To facilitate a fair comparison between OW and TW models, assuming that both transmission directions can utilize an ML-based encoder, we develop Active Feedback LightCode (ALC).\footnote{We constrain the transmit power of the feedback encoder so that it matches the power constraint for the encoder transmitting the message, i.e., if $P=1$ and $T=9$, the RHS of \eqref{eq:pwr_constraint} is $9$ for both the forward and feedback encoders.} To implement ALC, we use the TWLC encoder, with User 2 always fed $\bm{b}_2=\bm{0}$ as a dummy initialization; the decoder is only fed $[\bm{y}_2,\textcolor{black}{\bm{c}_2}]^{\intercal}$ since $\bm{b}_2$ provides no pertinent information. The overall loss is given by User 1's CE loss. The layer dimensions of ALC in Tab. \ref{tab:alc} are matched exactly to those in Figs. 12 \& 13 in \cite{ankireddy_lightcode_2024}; in Fig. \ref{fig:results_all}(a), its dimensions are matched to TWLC's. \textcolor{black}{We note here that, under an appropriate training setup, it should in principle be possible to realize an implementation of ALC by training TWLC, disabling user $2$'s decoder, and sending only $\bm{0}$ from user $2$, thereby allowing the system to adaptively select its transmit mode based on the need of user $2$ to transmit bits; this, however, adds to the complexity and size of the deployed model.}

\noindent\textbf{TW GBAF-Based System (TWBAF)}: Beginning with the encoder/decoder modules presented in \cite{ozfatura_gbaf_2022},  we insert \textcolor{black}{(with some abuse of notation) the function $\tanh{(\bm{X}-\bm{1}\mathbb{E}[\bm{X}]^{\intercal})}$ between the encoder's output and the power reallocation mechanism, where $\bm{X}\in\mathbb{R}^{B\times\ell}$ is the raw encoder output}, $B$ is the batch size, $\ell$ is the number of rows in the input matrix, and \textcolor{black}{$\mathbb{E}[\bm{X}]\in\mathbb{R}^{\ell}$} is taken over the elements in the batch. \textcolor{black}{This elementwise $\tanh$ function allows us to prevent divergent transmit symbols by mapping to the range $[-1,1]$ before power reweighting.} Bits are not BPSK modulated at the model's input. Without these updates, the TW model experienced training instability, failing to converge in all cases of $SNR_1=-1$ dB. \textcolor{black}{Unlike \cite{ozfatura_gbaf_2022}, we only utilize time-indexed power reallocation via weights $\bm{w}_i=[w_i^1,...,w_i^T]^{\intercal}$ rather than also reallocating along the block-dimension of each set of generated transmit symbols at time $t$; this permitted BLER improvement over reallocating along both dimensions.} We use feature extractor C with ReLU activations, while matching the original specifications of the dimensions of the encoder/decoder attention modules from \cite{ozfatura_gbaf_2022}. \textcolor{black}{We forego the belief unit from \cite{ozfatura_gbaf_2022} to keep the models comparably sized and reduce complexity. Despite having a similar FLOPS count to TWRNN (Sec. \ref{sec:res_flops}), TWBAF trained about $4\times$ faster, which may be beneficial for on-device training.}

\section{Results and Discussion}\label{sec:results}
In this section, we first assess the computational requirements for the encoders and decoders in terms of FLOPS. We then analyze the performance of TW versus OW coding under symmetric SNRs, and we conclude by comparing the TW models' efficacies across a variety of SNR regimes. \textcolor{black}{We set message lengths to $K_i=K$ and powers to $P_i=0$ dB, $\forall i$. For reference, we provide sum BLER results for $(T,K)$ polar codes (as in e.g., \cite{hkim_deepcode_2020}) over an open-loop TW channel.}\footnote{Code available: \url{https://github.com/DavidRNickel/twoway_models.git}.}

\subsection{Comparison of Floating-Point Operations 
(FLOPS)}\label{sec:res_flops}
We quantify the computational overhead by calculating the total number of FLOPS to process a single bitstream. Let $h_{\{c,b,r\}}$ be the size of a model's hidden dimensions, where $c,b,r$ denote TWLC, TWBAF, and TWRNN, resp.; $\ell$ be the number of rows in a TWBAF input matrix; $L_B^E$ and $L_B^D$ be the number of attention layers in TWBAF's encoder and decoder, resp.; and $K,~M,~T$ be the message-length, sub-message-length, and number of channel uses, resp., with $T_M=TM/K$. \textcolor{black}{We report the Big-$\mathcal{O}$ order of FLOPS for each method in Tab. \ref{tab:flops},\footnote{Full (handwritten) derivation located in our GitHub repository.} For subsequent experiments, we set $h_c=32$, $h_b=32$, and $h_r=50$, leaving approximately $0.6M$, $2.5M$, and $2.6M$ FLOPS for TWLC, TWBAF, and TWRNN, respectively.}


\begin{table}[t]
    \small
    \centering
    \caption{\small \textcolor{black}{Big-$\mco$} FLOPS comparison between models, where $K,~M,~T,~h_{\{c,b,r\}}$ are the message-length, sub-message-length, total number of channel uses, and hidden dimensions, resp. Equations are for a single user's encoder and decoder, i.e., half of the total FLOPS.}\label{tab:flops}
    \textcolor{black}{\begin{tabular}{|p{.45in}|p{3.17cm}|p{3.17cm}|}
    \hline
    $\textbf{Model}\downarrow$ & $\hspace{12mm}\textbf{Encoding}$ &\hspace{12mm}$\textbf{Decoding}$ \\
    \hline
    TWLC & \centering{$\mco (Th_c(M+T_M)+h_c^2)$} & \centering{$\mco(\sfrac{Kh_c}{M}(M+T_M)+h_c^2+h_c2^M)$}\tabularnewline
    \hline
    TWBAF & \centering{$\hspace{-1.2mm}\mco(T_M(L_B^E\ell h_B^2 + \ell h_B(T_M+M)))$} & \centering{$\mco(L_B^D\ell h_B^2+\ell h_B(M+T_M) + 2^{M+1}\ell h_B)$}\tabularnewline
    \hline
    TWRNN & \centering{$\mco(Th_r^2+Mh_r)$}  & \centering{$\mco(\sfrac{K}{M}(T_M(h_r^2+Mh_r)+2^{M}h_r))$} \tabularnewline
    \hline
    \end{tabular}}
\end{table}

\subsection{Comparison of One-way Versus Two-way Communications}\label{sec:res_ow_tw}
\textcolor{black}{In this section, we explore OW versus TW feedback communication. In Tab. \ref{tab:alc}, ALC shows considerable BLER improvement over LC, making it more competitive for comparison. We consider two cases: TWLC versus ALC, and the TW linear scheme (TWLIN) from \cite{jkim_rnntwc_2023} against its predecessor, the inner code of Chance and Love (CL) \cite{chance_linearcoding_2011}.\footnote{We consider the CL scheme with symbols from a 4-PAM constellation.} Users transmit $K$ bits apiece over a total of $T$ channel uses. For the OW case, each user is allotted $T/2$ channel uses, so $R_{ow}=2K/T$. For TW, the length-$K$ bitstreams are split into sub-blocks of length $M=K/2$; those $M$ bits are transmitted over $T/2$ channel uses, yielding $R_{tw}=K/T$. The decoders' outputs are then concatenated for BLER calculation. We consider settings of $K=6,~T=18$ and $K=4,~T=16$.}

\textcolor{black}{Looking first to the ML-based models, we find in Fig. \ref{fig:results_all}(a) distinct regions in which OW and TW communications are advantageous. For $K=6$, the lower coding rate enabled by TW coding admits uniformly lower BLER. However for $K=4$, the OW model generally performs better. Turning to the linear schemes, the $K=6$ data seems to reverse the BLER/SNR trend from TWLC/ALC. We may attribute these results to the optimization scheme of TWLIN, which we found not particularly robust. For $K=4$, TWLIN shows uniform improvement over CL, again showing benefit in the lower coding rate permitted by TW schemes. These results suggest that while there is great potential benefit in utilizing TW schemes, it is worthwhile to benchmark against a comparable OW scheme when deploying a model to ensure a desirable BLER/complexity tradeoff.}

\begin{table}[t]
    \centering
    \small
    \caption{\small BLER performance of one-way Active LightCode vs standard LightCode for $SNR_2=20$ with $K=M=3$ and $T=9$.}\label{tab:alc}
    \begin{tabular}{|c|c|c|c|}
    \hline
    $\mathbf{SNR_1}\text{[dB]}\rightarrow$ & $\mathbf{-1}$ & $\mathbf{0}$ & $\mathbf{1}$ \\
    \hline
    \textbf{BLER (LC)} & $3.9\mathrm{e}\hspace{.4mm}\text{--}\hspace{.4mm}3$ & $8.4\mathrm{e}\hspace{.4mm}\text{--}\hspace{.4mm}5$ & $2.1\mathrm{e}\hspace{.4mm}\text{--}\hspace{.4mm}7$\\
    \hline
    \textbf{BLER (ALC)} & \textcolor{black}{$9.8\mathrm{e}\hspace{.4mm}\text{--}\hspace{.4mm}6$} & \textcolor{black}{$3.0\mathrm{e}\hspace{.4mm}\text{--}\hspace{.4mm}8$} & \textcolor{black}{$4.0\mathrm{e}\hspace{.4mm}\text{--}\hspace{.4mm}9$}\\
    \hline
    \end{tabular}
\end{table}

\subsection{Two-Way BLER Comparison}\label{ssec:results_shortblock}
\begin{figure*}[t]
    \centering
    \includegraphics[width=0.9\textwidth]{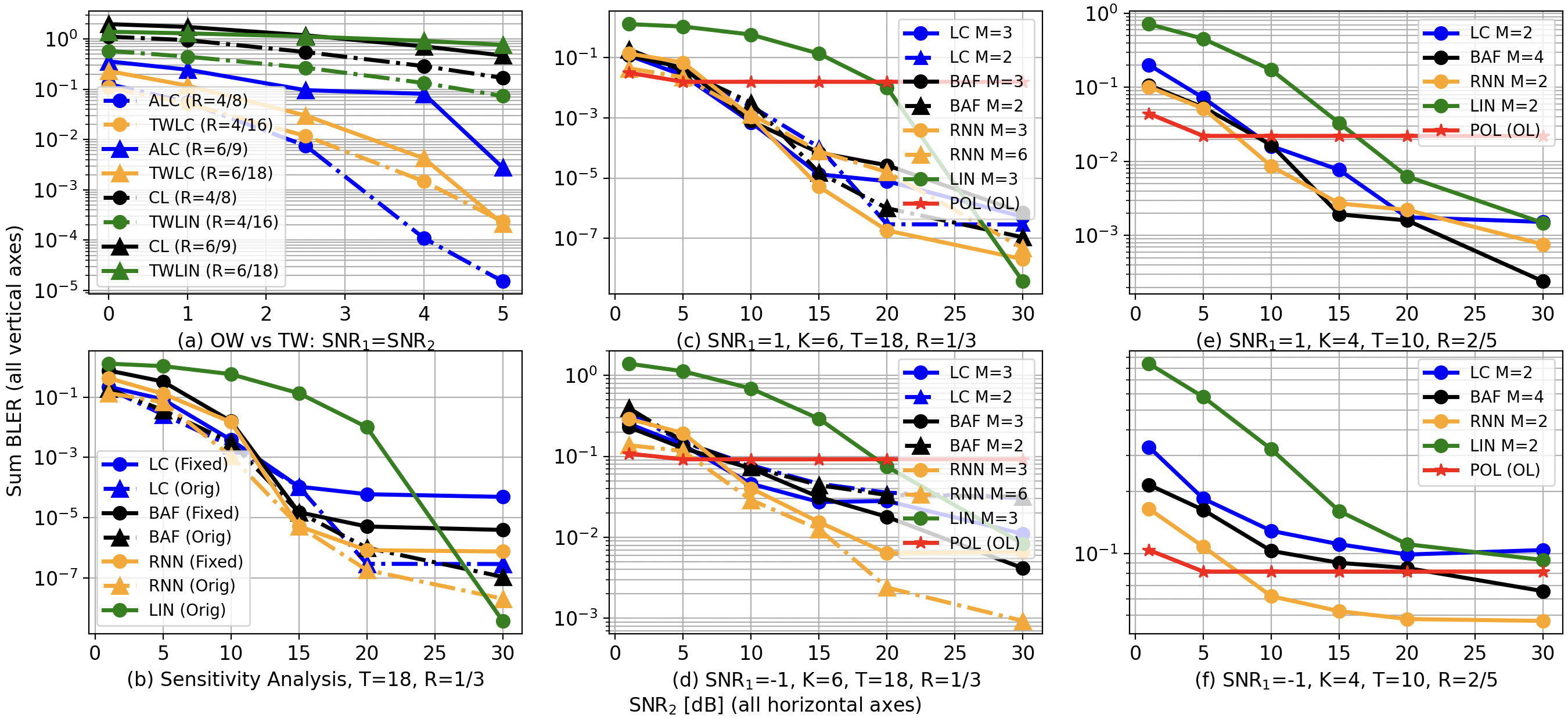}
    \caption{\small \textcolor{black}{Sum BLER results for all experiments. ``TW" is truncated from the model names in legends in Figs. (b-f) for brevity. In (a), we compare OW and TW communications transmitting the same total number of bits over the same total channel uses (yielding rate $R$, as explained in Sec. \ref{sec:res_ow_tw}), finding distinct regions over which each is more advantageous. (b) shows the sensitivity of each model to out-of-training-distribution noise, finding that matching noise between training and inference can play a substantial role in performance. (c,d) and (e,f) explore rate $1/3$ and $2/5$ codes resp., with $K$ denoting the total number of bits, $T$ the block-length, and $M$ the ``sub-block" size into which a length-$K$ bitstream is split for encoding. ``POL (OL)" denotes an open-loop TW channel with each user using a $(T,K)$ polar code.}}\label{fig:results_all}
\end{figure*}

In Fig. \ref{fig:results_all} (c-f), we explore rate $R$ codes on messages of length $K$. For length-$M$ sub-blocks, $TM/K$ channel uses apiece are utilized in transmitting $K/M$ sub-blocks whose results are concatenated for BLER calculation of the length-$K$ block. In Fig. \ref{fig:results_all}(c), all models perform comparably in lower $SNR_2$ regimes, making TWLC, with its smaller architecture, a viable option. TWRNN and TWBAF perform noticeably better as $SNR_2$ increases, indicating their ability to capture more complex relationships between the users. \textcolor{black}{For $SNR_1=-1$ dB in Fig. \ref{fig:results_all}(d,f), we find that TWRNN clearly outperforms all other models. The performance trends seen in the rate $1/3$ setting continue into the rate $2/5$ regime of Fig. \ref{fig:results_all}(e,f), with the TWLIN demonstrating mostly inferior BLER to the ML-based models. While we expect model performance would suffer at higher coding rates, this is not the job of these physical layer models; such higher coding rates are achieved via outer codes.}

Lastly in Fig. \ref{fig:results_all}(b), the sensitivity of each model is assessed against out-of-distribution (OOD) noise by first training a ``fixed" model on $SNR_1=1~\text{dB},~SNR_2=15$ dB then testing the model as a function of $SNR_2$. While all models are reasonably robust to OOD noise, there is distinct benefit in tailoring the training to closely match the intended use-case, as evidenced by the superiority of the lines labeled ``Orig", run with matched train/test SNR. \textcolor{black}{We also observe that TWLC trained under fixed $SNR_2$ has an error floor that is noticeably higher than the other models, reached around $SNR_2 = 15$ dB. This indicates that the more intricate RNN and attention mechanism structures employed by TWRNN and TWBAF may enable better generalizability to settings with favorable noise conditions during inference relative to training.}

\subsection{Key Takeaways}
TWLC, TWBAF, and TWRNN all demonstrate viability under different engineering goals and system settings. From Sec. \ref{sec:res_ow_tw}, we find that one-way and two-way communications demonstrate distinct SNR regions and coding rates over which they respectively perform better. Examining the distinct models, we find that TWLC obtains decent performance with a significantly smaller network, requiring under $25\%$ of the FLOPS to run as the other two (Sec. \ref{sec:res_flops}). However, TWLC is relatively less capable of capturing the intricate patterns available in certain SNR regimes. TWRNN demonstrates superior BLER performance in low $SNR_1$ regimes but suffers from large size and computational cost due to its recurrent structure. Lastly, TWBAF performs comparably to TWRNN in higher $SNR_1$ settings, \textcolor{black}{making it a viable alternative}. The high performance of TWRNN and TWBAF appears in large part to be due to their utilization of attention mechanisms.

\section{Conclusions}\label{sec:conc}
In this paper, we presented an algorithmic framework for ML-based two-way feedback coding. We then explored the performance tradeoffs of one-way and two-way communications. We find that in some noise regimes, two-way coding demonstrates superior performance to comparable one-way systems. Lastly, we explored the efficacy of three two-way models, finding that each is meritorious depending on the desired engineering outcomes. These efforts can guide future endeavors in the development of novel coding schemes, adaptive feedback mechanisms, and performance optimization. 

\bibliography{references}

\begin{thebibliography}{10}
\providecommand{\url}[1]{#1}
\csname url@samestyle\endcsname
\providecommand{\newblock}{\relax}
\providecommand{\bibinfo}[2]{#2}
\providecommand{\BIBentrySTDinterwordspacing}{\spaceskip=0pt\relax}
\providecommand{\BIBentryALTinterwordstretchfactor}{4}
\providecommand{\BIBentryALTinterwordspacing}{\spaceskip=\fontdimen2\font plus
\BIBentryALTinterwordstretchfactor\fontdimen3\font minus \fontdimen4\font\relax}
\providecommand{\BIBforeignlanguage}[2]{{%
\expandafter\ifx\csname l@#1\endcsname\relax
\typeout{** WARNING: IEEEtran.bst: No hyphenation pattern has been}%
\typeout{** loaded for the language `#1'. Using the pattern for}%
\typeout{** the default language instead.}%
\else
\language=\csname l@#1\endcsname
\fi
#2}}
\providecommand{\BIBdecl}{\relax}
\BIBdecl

\bibitem{shannon_twc_1961}
C.~E. Shannon, ``Two-way communication channels,'' \emph{Proceedings of the Fourth Berkeley Symposium on Mathematical Statistics Probability, Vol. 1: Contributions to the Theory of Statistics}, pp. 611--644, Jan. 1961.

\bibitem{han_independentgtwc_1984}
T.~Han, ``A general coding scheme for the two-way channel,'' \emph{IEEE Trans. Info. Theory}, vol.~30, no.~1, pp. 35--44, 1984.

\bibitem{palbaus_errorexponent_2021}
K.~S. Palacio-Baus and N.~Devroye, ``Achievable error exponents of one-way and two-way {AWGN} channels,'' \emph{IEEE Trans. Info. Theory}, vol.~67, no.~5, pp. 2693--2715, 2021.

\bibitem{sk_gowc_1966}
J.~Schalkwijk, ``A coding scheme for additive noise channels with feedback--{II}: Band-limited signals,'' \emph{IEEE Trans. Info. Theory}, vol.~12, no.~2, pp. 183--189, 1966.

\bibitem{chance_linearcoding_2011}
Z.~Chance and D.~J. Love, ``Concatenated coding for the {AWGN} channel with noisy feedback,'' \emph{IEEE Trans. Info. Theory}, vol.~57, no.~10, pp. 6633--6649, 2011.

\bibitem{yishai_fbkcode_2017}
A.~Ben-Yishai and O.~Shayevitz, ``Interactive schemes for the awgn channel with noisy feedback,'' \emph{IEEE Trans. on Info. Theory}, vol.~63, no.~4, pp. 2409--2427, 2017.

\bibitem{hkim_deepcode_2020}
H.~Kim, Y.~Jiang, S.~Kannan, S.~Oh, and P.~Viswanath, ``Deepcode: Feedback codes via deep learning,'' \emph{IEEE Jour. Sel. Area. Info. Theory}, vol.~1, no.~1, pp. 194--206, 2020.

\bibitem{vaswani_xformer_2017}
A.~Vaswani, N.~Shazeer, N.~Parmar, J.~Uszkoreit, L.~Jones, A.~N. Gomez, L.~Kaiser, and I.~Polosukhin, ``Attention is all you need,'' \emph{Advances in neural information processing systems}, vol.~30, 2017.

\bibitem{ozfatura_gbaf_2022}
E.~Ozfatura, Y.~Shao, A.~G. Perotti, B.~M. Popović, and D.~Gündüz, ``All you need is feedback: Communication with block attention feedback codes,'' \emph{IEEE Jour. Sel. Area. Info. Theory}, vol.~3, pp. 587--602, 2022.

\bibitem{shao_attentioncode_2023}
Y.~Shao, E.~Ozfatura, A.~G. Perotti, B.~M. Popović, and D.~Gündüz, ``{AttentionCode}: Ultra-reliable feedback codes for short-packet communications,'' \emph{IEEE Trans. Commun.}, vol.~71, no.~8, pp. 4437--4452, 2023.

\bibitem{ozfatura_baaf_2023}
E.~Ozfatura, Y.~Shao, A.~Ghazanfari, A.~Perotti, B.~Popovic, and D.~Gündüz, ``Feedback is good, active feedback is better: Block attention active feedback codes,'' in \emph{IEEE International Conference on Communications}, 2023, pp. 6652--6657.

\bibitem{ankireddy_lightcode_2024}
S.~K. Ankireddy, K.~R. Narayanan, and H.~Kim, ``{LightCode}: Light analytical and neural codes for channels with feedback,'' \emph{IEEE Jour. Sel. Area. Commun.}, vol.~43, no.~4, pp. 1230--1245, 2025.

\bibitem{jkim_robustcode_2023}
J.~Kim, T.~Kim, D.~J. Love, and C.~Brinton, ``Robust non-linear feedback coding via power-constrained deep learning,'' in \emph{International Conference on Machine Learning}.\hskip 1em plus 0.5em minus 0.4em\relax PMLR, 2023, pp. 16\,599--16\,618.

\bibitem{malayter2024deeplearningaidedbroadcast}
\BIBentryALTinterwordspacing
J.~Malayter, C.~Brinton, and D.~J. Love, ``Deep learning aided broadcast codes with feedback,'' 2024. [Online]. Available: \url{https://arxiv.org/abs/2410.17404}
\BIBentrySTDinterwordspacing

\bibitem{zhou2025learnedcodesbroadcastchannels}
\BIBentryALTinterwordspacing
Y.~Zhou and N.~Devroye, ``Learned codes for broadcast channels with feedback,'' 2025. [Online]. Available: \url{https://arxiv.org/abs/2411.04083}
\BIBentrySTDinterwordspacing

\bibitem{jkim_rnntwc_2023}
\BIBentryALTinterwordspacing
J.~Kim, T.~Kim, A.~B. Das, S.~Hosseinalipour, D.~J. Love, and C.~G. Brinton, ``Coding for gaussian two-way channels: Linear and learning-based approaches,'' \emph{Accepted to IEEE Trans. Info. Theory}, 2025. [Online]. Available: \url{https://arxiv.org/abs/2401.00477}
\BIBentrySTDinterwordspacing

\end{thebibliography}
\bibliographystyle{IEEEtran}
\end{document}